\documentclass[cite,figures]{epl}

\usepackage{graphicx}
\begin{document}
\title{
Breakdown of the Luttinger sum-rule at the Mott-Hubbard transition
in the one-dimensional $t_1-t_2$ Hubbard model.}
\shorttitle{Breakdown of the Luttinger sum-rule at the Mott-Hubbard transition}

\author{Claudius Gros\inst{1},Kay Hamacher\inst{2} and
         Wolfgang Wenzel\inst{3}}
\institute{
\inst{1} 
Fakult\"at 7, Theoretische Physik,
University of the Saarland, 66041 Saarbr\"ucken, Germany.\\
\inst{2}
Dept.~of Physics, University of California, San Diego,
9500 Gilman Drive, MC0374, La Jolla, CA 92093, USA.\\
\inst{3}
Forschungszentrum Karlsruhe,
Institut f\"ur Nanotechnologie,
Hermann-von-Helmholtz-Platz 1,
76344 Eggenstein-Leopoldshafen, Germany.
          }

%\pacs{ 75.30.Gw, 75.10.Jm, 78.30.-j }
\maketitle

\begin{abstract}
We investigate the momentum distribution function near the
Mott-Hubbard transition in the one-dimensional $t_1-t_2$ Hubbard model
(the zig-zag Hubbard chain), with the density-matrix
renormalization-group technique. 
We show that for strong interactions
the Mott-Hubbard transition occurs between the metallic-phase and an
insulating dimerized phase with incommensurate
spin excitations, suggesting a decoupling of magnetic and charge
excitations not present in weak coupling. 
We illustrate the signatures
for the Mott-Hubbard transition and the commensurate-incommensurate
transition in the insulating spin-gaped state in their respective
ground-state momentum distribution functions.
\end{abstract}

{\bf Introduction -} 
Many aspects of the low-energy physics of an electronic system are
influenced by the shape of its Fermi surface and the occupation of
nearby states. Finite temperature, disorder or strong correlations may
change the Fermi-surface geometry and or topology and induce magnetic
and other instabilities. 

%% KH 
The investigation of the interplay of magnetic interactions, in
particular the existence of incommensurate phases, with charge
excitations emerges as one of the central issues in the physics of
low-dimensional electronic 
systems~\cite{shraiman89,sarker91,zaanen89,poilblanc89,white00,jec02,san04}.  
The nature and mechanism of the
Mott-Hubbard transition, as one of its most striking manifestations,
has been subject to intense investigation for many years.  It is
therefore important to study the renormalization of individual
Fermi-surface sections under the influence of electron-electron
correlations in competition with frustrating interactions.

Recent investigations suggested a
spontaneous, interaction induced deformation of the Fermi surface of
the 2D (extended) Hubbard model close to half filling, indicating in
part a violation of the Luttinger
sum-rule~\cite{gro94,yod99,val00}. The possibility of a relevant
Fermi-surface renormalization near the Mott-Hubbard (MH) transition in
one-dimensional generalized Hubbard models has been raised on the
basis of a DMRG-study~\cite{ham02}, drawing on arguments from results
obtained for the Fermi-surface flow by RG~\cite{lou01,dus03}.

In this letter we examine this Mott-Hubbard transition in one of the
prototypical frustrated one-dimensional models: the half-filled
zig-zag Hubbard ladder. For appropriate choices of the parameters this
model describes either the low-energy properties of
Hubbard-ladders~\cite{lou01} or half-filled edge-sharing double-chain
materials like SrCuO$_2$~\cite{ric93} or LiV$_2$O$_5$~\cite{val01} for
which the next-nearest neighbor hopping is expected to be
substantial. The study of this system permits both the consideration
of incommensurate phases and strong interactions. The former are
difficult to study in the limit of infinite dimensions, where part of our
present understanding of the Mott-Hubbard transition orginates. The
latter can now be treated adequately for one-dimensional systems using
the density-matrix renormalization group (DMRG), which emerged in the
last decade as a reliable tool to investigate the electronic structure
of quasi one-dimensional systems \cite{dmrgpaper,dmrgbuch}.

We find no renormalization to
perfect-nesting of the Fermi-surface in the metallic state as the 
Mott-Hubbard transition is approached.
We explicitly demonstrate that the dimerized state on the insulating side
of the Mott-Hubbard transition has gaped incommensurate
spin-excitations, which only later give way to an insulating phase
with commensurate spin-excitations. Our results indicate that
perfect-nesting is not a prerequisite for the Mott-Hubbard transition
at finite values of the interaction. The opening of the charge gap
decouples from changes in the nature of magnetic excitations, in stark
contrast to the weak-coupling scenario, where Umklapp-scattering
becomes relevant at perfect nesting and leads to the opening of a
charge-gap. We report the characteristic signatures of the
incommensurate insulating state in momentum-distribution function for
future experimental analysis.
% ------------------ %
% ------------------ %
% ------------------ %
% ------------------ %
%%%%%%%%%%%%%%%%%%%%%%%%%%%%%%%%%%%%%%%%%%%%%%%%%%%%%%%%%%%%%%%%%%%
%%%%%%%%%%%%%%%%%%%%%%%%%%%%%%%%%%%%%%%%%%%%%%%%%%%%%%%%%%%%%%%%%%%

{\bf Model} - 
The Hamiltonian of the zig-zag Hubbard ladder
is given as:
\begin{equation}
H  =
\sum_{n,\sigma\atop\Delta n=1,2}
t_{\Delta n}
\left(
c_{n+\Delta n,\sigma}^\dagger c_{n,\sigma}^{\phantom{\dagger}}
+ {\rm h.c.}\right)  +\, U\sum_{n} n_{\uparrow}n_{\downarrow},
% &&\quad\qquad \,+\, U\sum_{n} n_{\uparrow}n_{\downarrow}
%c_{n,\uparrow}^\dagger c_{n,\uparrow}^{\phantom{\dagger}}
%c_{n,\downarrow}^\dagger c_{n,\downarrow}^{\phantom{\dagger}}~,
\label{def_H}
\end{equation}
where the $c_{n,\sigma}^\dagger$ ($c_{n,\sigma}^{\phantom{\dagger}}$)
are Fermion-creation (destruction) operators on site $n$ and spin
$\sigma=\uparrow,\downarrow$ and $n_{\sigma}=c_{n,\sigma}^\dagger c_{n,\sigma}^{\phantom{\dagger}}$. 

\begin{figure}[t]
\centerline{\includegraphics[width=0.7\columnwidth]{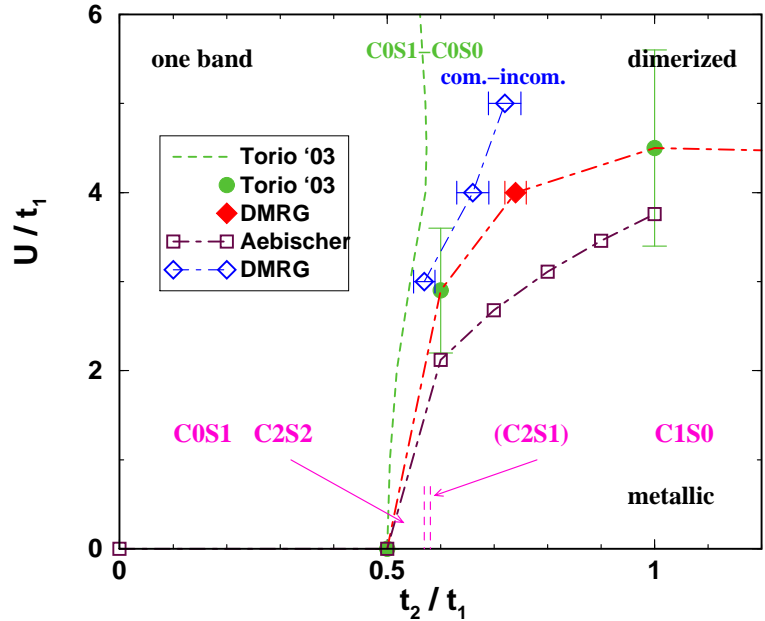}}
\smallskip
\caption{\label{fig_phaseDia}
Phase diagram for the half-filled $t_1-t_2$ Hubbard
model as explained in the text. Included are the
DMRG-result for the Mott-Hubbard transition at $U=4t_1$
(filled diamonds) and for the commensurate-incommensurate
transition in the insulating state (open diamonds). The
dashed-dotted lines are guides to the eye. 
The states $CnSm$ predicted by weak-coupling RG are denoted
at the bottom of the phase diagram.
         }
\end{figure}

Let us first discuss a few known properties of the phase-diagram
at half filling, compare Fig.\ \ref{fig_phaseDia}.
We follow
Balents and Fisher\cite{bal96} and denote
with
\[
CnSm
\]
a phase with $n/m$ {\em gapless} charge/spin modes. 
A Mott-Hubbard transition of type
C0S1--C2S2 is predicted \cite{lou01} by RG
at half-filling
for $t_2=t_1/2$. This prediction should be valid at
infinitesimal $U/t_1$.

In the limit of large $U$ the model (\ref{def_H})
transforms to the $J_1-J_2$ chain
\begin{equation}
H_J = \sum_{n,\sigma\atop\Delta n=1,2}
J_{\Delta n}\, {\bf S}_n\cdot {\bf S}_{n+\Delta n}~,
\label{def_H_J}
\end{equation}
with $J_{\Delta n}=4t_{\Delta n}^2/U$.
The $J_1-J_2$ model spontaneously dimerizes\cite{Hal82}
for big enough $J_2$. The critical value for
$\alpha = J_2/J_1$ can be determined by examination
of the nature of the lowest-lying excitation for
finite clusters\cite{Nom94}.
In the spin-fluid state C0S1 it is a triplet,
in the dimerized state C0S0 it is a singlet (the
dimerized state is doubly degenerate in the
thermodynamic limit). One finds\cite{Nom94}:
$\alpha_c=0.2411=(0.491)^2$.

Torio {\it et al.}\cite{tor03} 
have extended the level-crossing
study used to determine $\alpha_c$ for the $J_1-J_2$
model to finite $U/t_1$ see the dashed line in
Fig.\ \ref{fig_phaseDia}. They propose that the dimerization
line C0S1-C0S0 extends until $U/t_1\to0$ and connects
to $t_2=t_1/2$.  This would consequently invalidate
the weak-coupling RG-prediction of a Mott-Hubbard transition
of type C0S1-C2S2. It seems, presently, more likely that
the transition is of C0S0-CnSm type,
with $n>0$. 

The Majumbdar-Gosh point  $J_2=J_1/2$, also called
disorder-point, has an exact valence-bond
dimer ground-state.
For $\alpha>0.5=(0.707)^2$ the short range
spin-spin correlations become incommensurable. Due
to the absence of long-range order\cite{Sch96}, the peak
in the static structure factor $S(q)$ moves away
from $q=\pi$ only at the Lifschitz-point, which can
be determined via DMRG\cite{Bur95} as
$\alpha_L = 0.52063=(0.7215)^2$.

The classical spin-wave solution to (\ref{def_H_J}) yields a
long-ranged spiral ground-state for $\alpha>0.5$ \cite{Rao97,Whi96}. 
A long-ranged ordered states is unstable towards quantum-fluctuation
in 1D, they become short-ranged. For the $J_1-J_2$ model the
quantum fluctuations lead to a dimerized state with finite
dimerization and short ranged spiral (incommensurate) spin-spin 
correlations. They show up as a peak in $S(q)$ for a
incommensurate wavevector $q$ and width $1/\xi$, where
$\xi$ is the correlations length.
There has been no study so far of the extension
of the dimerized phase with incommensurate excitations, 
realized for $\alpha>0.5$ in the limit of large $U/t_1$ 
to finite values of $U/t_1$. Here we propose that it connects
to the metallic state.

Also included in Fig.\ \ref{fig_phaseDia} are the predictions
for the critical $U_c$ for the Mott-Hubbard transition
by Torio {\it et al.}\cite{tor03} and by 
Aebischer {\it et al.}\cite{aeb01}. It has turned out
to be very difficult to determine numerically
the location of this transition
from estimates of the charge gap, extrapolated to 
the thermodynamic limit, due to the fact that the charge gap is
exponentially small near the Mott-Hubbard transition.
Here we find that a quite accurate lower bound for the 
the Mott-Hubbard transitions can be obtained form the study
of the momentum distribution function.

%%%%%%%%%%%%%%%%%%%%%%%%%%%%%%%%%%%%%%%%%%%%%%%%%%%%%%%%%%%%%%%%%%%
%%%%%%%%%%%%%%%%%%%%%%%%%%%%%%%%%%%%%%%%%%%%%%%%%%%%%%%%%%%%%%%%%%%

% ------------------ %
% ------------------ %
\begin{figure}[t]
\centerline{\includegraphics[width=0.7\columnwidth]{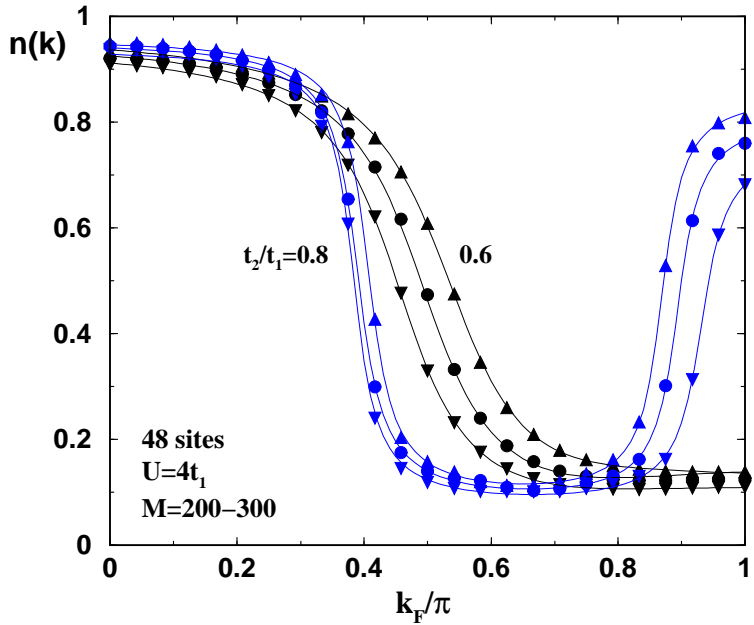}}
\smallskip
\caption{\label{fig_nk_fit}
Illustration of the momentum-distribution function
$n(k)$ for $L=48$ sites, $U=4t_1$ and 
$t_2/t_1=0.6,0.8$. DMRG results for
$N_\uparrow=N_\downarrow=L/2$ (filled circles) and for
$N_\uparrow=L/2+2$, $N_\downarrow=L/2-2$ 
(triangles-up: majority-spin, triangles-down: minority-spin).
The lines are fits by Eq.\ (\ref{fit_nk}).
         }
\end{figure}
{\bf Method} - In the last decade the density matrix renormalization
group~\cite{dmrgpaper} emerged as a reliable tool for the study
of electronically one-dimensional models~\cite{dmrgbuch}. For strongly
interacting fermionic systems, the evaluation of $n(k)$ as the Fourier
transform of the correlation function:
\begin{equation}
n_\sigma(k)\ =\ {2\over L}\sum_{n,n'=1}^L
\cos(k(n-n'))\, \langle\,
c_{n,\sigma}^\dagger
c_{n',\sigma}^{\phantom{\dagger}}\, \rangle
\label{nk_real}
\end{equation}
is numerically difficult and costly in the framework of the
DMRG~\cite{dmrgpaper,dmrgbuch}, in particular for periodic boundary
conditions~\cite{qin95}. 
Here we report results from DMRG calculations on half-filled
chains of length L=48,~80 using up-to 300 DMRG states, 
which we found sufficiently accurate to determine $n(k)$.
% ------------------ %
% ------------------ %
% ------------------ %
% ------------------ %
\begin{figure}[t]
%\noindent
%\\
\centerline{
\includegraphics[width=0.7\textwidth,angle=0]{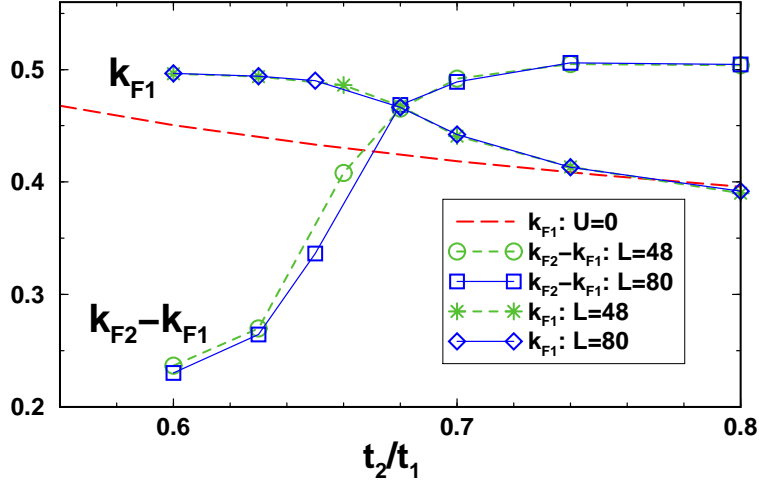}
}
\smallskip
\caption{\label{fig_kf_shifts}
For $U=4t_1$ and $L=48,80$ sites the DMRG
results for $k_{F1}$ and the difference
$k_{F2}-k_{F1}$ (both in units of $\pi$)
as obtained by fitting
the DMRG-results for $n(k)$ by Eq.\ (\ref{fit_nk}),
compare Fig.\ \ref{fig_nk_fit}. 
The lines are guides to the eye.
Below $t_2/t_1\approx0.74$
two things happen: $k_{F1}$ renormalizes substantially
with respect to its $U=0$-value (dashed line) and 
$k_{F2}-k_{F1}$ deviated from its Luttinger-sum-rule-value 
of $0.5$. 
}
\end{figure}
% ------------------ %
% ------------------ %

%%%%%%%%%%%%%%%%%%%%%%%%%%%%%%%%%%%%%%%%%%%%%%%%%%%%%%%%%%%%%%%%%%%
%%%%%%%%%%%%%%%%%%%%%%%%%%%%%%%%%%%%%%%%%%%%%%%%%%%%%%%%%%%%%%%%%%%
To obtain estimates for the Fermi wavevectors we have analyzed the
momentum distribution function, see Eq.\ (\ref{nk_real}), obtained by
DMRG, via two smoothed step functions at $k_{F1}$ and $k_{F2}$ with
respective widths $p_1$ and $p_2$ \cite{ham02}:
\begin{equation}
n(k)\ =\ a_0 + a_1\,{\rm atan}{k-k_{F1}\over p_1}
             + a_2\,{\rm atan}{k-k_{F2}\over p_2}~.
\label{fit_nk}
\end{equation}
The quality of fits by (\ref{fit_nk}) to $n(k)$ 
is illustrated in Fig.\ \ref{fig_nk_fit}. We will use the
such obtained estimates for the Fermi-wavevectors
$k_{F1}$ and $k_{F2}$ in the further analysis.

{\bf Results} - In order to elucidate the properties of the Mott-Hubbard transition
\cite{aeb01} and of the incommensurate-commensurate transition in one
dimensional frustrated systems, we studied the 
momentum-distribution function for a range of parameters that crosses
all relevant phase transition-lines of the phase diagram of the 1D
$t_1-t_2$ Hubbard model~\cite{dau00}. In Fig.\ \ref{fig_kf_shifts} we
present our estimates for $k_{F1}$ as a function of $t_2/t_1$ for
$U=4t_1$, in comparison with the result for $U=0$. We note a
substantial renormalization of $k_{F1}$ towards larger values below
$t_2/t_1\approx 0.74$.

% ------------------ %
% ------------------ %
\begin{figure}[t]
\centerline{\includegraphics[width=0.7\columnwidth]{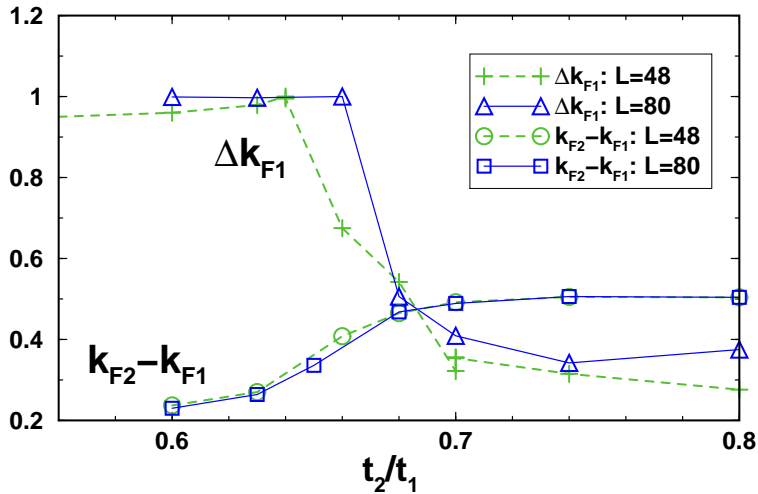}}
\smallskip
\caption{\label{fig_kf_delta}
The normalized shift 
$\Delta k_{F1}$, see Eq.\ (\ref{Delta_kF1})
for $L=48,80$ sites and $U=4t_1$. The
lines are guides to the eye. For comparison,
we have reproduced the results for $k_{F2}-k_{F1}$
from Fig.\ \ref{fig_kf_shifts}.
         }
\end{figure}
% ------------------ %
% ------------------ %

In the metallic state the Luttinger sum-rule states that the total
volume of both Fermi-seas together equals the number of electrons.
At half-filling this statement is equivalent with $k_{F2}-k_{F1}=\pi/2$.
Our DMRG-data for $\left(k_{F2}-k_{F1}\right)/\pi$ presented in Fig.\
\ref{fig_kf_shifts} indicates a violation of the Luttinger sum-rule
below $t_2/t_1\approx 0.74$. We therefore conclude that 
the Mott-Hubbard transition occurs in the vicinity of
this point (as indicated in Fig.\
\ref{fig_phaseDia} by the filled diamond 
in Fig. \ref{fig_phaseDia}) and that the system is
insulating for $t_2/t_1<0.74$. 
Due to the exponentially small gap in this region our result
for the position of the Mott-Hubbard transition is in fact
a lower bound (in terms of $t_2/t_1$) for the exact transition
point, as we would not be able to resolve numerically a possible
exponentionally small departure from the Luttinger-liquid
sum-rule for $t_2/t_1>0.74$.

%%%%%%%%%%%%%%%%%%%%%%%%%%%%%%%%%%%%%%%%%%%%%%%%%%%%%%%%%%%%%%%%%%%
%%%%%%%%%%%%%%%%%%%%%%%%%%%%%%%%%%%%%%%%%%%%%%%%%%%%%%%%%%%%%%%%%%%
Next we turn to an analysis of the nature of the spin excitation in
the vicinity of the Mott-Hubbard transition.  The spin-excitation in
the insulating state correspond to renormalized particle-hole
excitations with a spin-flip.  They correspond therefore to states
with $N_\uparrow = L/2+\Delta N/2$ and $N_\downarrow = L/2-\Delta
N/2$.  Using for numerical convenience $\Delta N=2$
we have calculated the respective Fermi-wavevectors for the 
majority $k_{F1,\uparrow}$ 
and minority $k_{F1,\downarrow}$
spins.  We define the
normalized shift for the Fermi-wavevector as
\begin{equation}
\Delta k_{F1}\ =\ 
\left( k_{F1,\uparrow}-k_{F1,\downarrow}\right) L/(\pi\Delta N)~.
\label{Delta_kF1}
\end{equation}
The normalization in Eq.\ (\ref{Delta_kF1}) is chosen
such that $\Delta k_{F1}=1$, independent of system size $L$,
for case of a single Fermi-sea. In the limit
$t_2/t_1\to\infty$ the Hubbard zig-zag chain has two
equally large Fermi-seas, with equal Fermi-velocities,
and $\Delta k_{F1}\to0.5$ in this limit, as we have
verified numerically for large $t_2/t_1$.

We have obtained $\Delta k_{F1}$
by fitting the respective momentum
distribution functions for the majority and minority spins,
see Fig.\ \ref{fig_nk_fit}, by Eq.\ (\ref{fit_nk}).
The results are presented in Fig.\ \ref{fig_kf_delta}.
We notice that $\Delta k_{F1}\approx 1$ for
small $t_2/t_1$, implying commensurated spin excitations.
A well defined kink in
$\Delta k_{F1}$ at $t_2/t_1\approx0.64-0.66$ indicates
a second-order transition to a state with incommensurate
spin-excitations (denoted by the open diamond in
Fig.\ \ref{fig_phaseDia}).

These results imply that the Mott-Hubbard transition takes
place for Fermi-wavevectors at arbitrary, incommensurate values, and
the Fermi-wavevectors do not renormalize towards perfect nesting
$k_{F1}\to \pi/2$ and $k_{F2}\to \pi$. In weak-coupling,
Umklapp-scattering becomes relevant at perfect nesting and leads to
the opening of a charge-gap. Our results indicate, that effective
perfect-nesting is not necessary for the Mott-Hubbard transition for
finite values of the interaction-strength.

%%%%%%%%%%%%%%%%%%%%%%%%%%%%%%%%%%%%%%%%%%%%%%%%%%%%%%%%%%%%%%%%%%%
%%%%%%%%%%%%%%%%%%%%%%%%%%%%%%%%%%%%%%%%%%%%%%%%%%%%%%%%%%%%%%%%%%%
{\bf Discussion} - The interplay between magnetic order and carrier
mobility in strongly interacting systems, with the Mott-Hubbard
transition as one important manifestation, has long been subject to
intense investigation. For the two-dimensional (2D) case, important for
experimental realizations, exact analytical or
numerical techniques are still lacking. It is known that in 2D the
commensurate ordered phase for the nearest-neighbor Hubbard model is
unstable with respect to helical
fluctuations~\cite{shraiman89,sarker91} either upon doping or in the
presence of frustrating next-nearest-neighbor interactions. Other,
even more exotic phases~\cite{zaanen89,poilblanc89} have also been
proposed, in particular in the context of studies into the mechanism
of high-temperature superconductivity. Our investigation of the
one-dimensional case offers new insight into driving forces behind the
MH transitions: nesting, i.e. the divergence of the magnetic
susceptibility due to special features of the Fermi surface geometry
emerges as sufficient, but apparently not a necessary condition of the
MH transition~\cite{jay91}. This result is inaccessible by weak
coupling theory and indicates the existence of a generic mechanism for
charge carrier freezing that is independent of dominant Umklapp
scattering processes. Spin-charge separation in Luttinger liquids may
enhance this effect in one-dimension, but similar
strong-coupling scenarios must be present also in higher dimensions to
explain the MH transition between incommensurate insulators and metals
in doped or frustrated 2D systems. 

%    ------------------
%    ------------------

{\bf Acknowledgments} -
We thank 
%% R.~Valent{\'\i} for a careful reading of the manuscript 
%%KH
%and 
D.~Baeriswyl and G.I.~Japaridze for stimulating discussions. We
thank the Deutsche Forschungsgemeinschaft, the BMBF and the
von-Neumann-Center for Scientific Computation for financial support.
KH is supported through a Liebig-Fellowship of the Fonds der
chemischen Industrie.

%%%%%%%%%%%%%%%%%%%%%%%%%%%%%%%%%%%%%%%%%%%%%%%%%%%%%%%%%%%%%5
%%%%%%%%%%%%%%%%%%%%%%%%%%%%%%%%%%%%%%%%%%%%%%%%%%%%%%%%%%%%%5

%%%%%%%%%%%%%%%%%%%%%%%%%%%%%%%%%%%%%%%%%%%%%%%%%%%%%%%%%%%%%5
%%%%%%%%%%%%%%%%%%%%%%%%%%%%%%%%%%%%%%%%%%%%%%%%%%%%%%%%%%%%%5

\end{document}